\documentstyle[multicol,aps]{revtex}
\begin{document}
\title{Quantum Phenomena in Low-Dimensional Systems} 
\author{Michael R. Geller}
\address{Department of Physics and Astronomy, University of Georgia, Athens, Georgia 30602-2451}
\date{\today}\maketitle

\begin{abstract}
A brief summary of the physics of low-dimensional quantum systems 
is given. The material should be accessible to advanced physics 
undergraduate students. References to recent review articles and 
books are provided when possible.
\end{abstract}
\begin{multicols}{2}

\section{introduction} 
\label{introduction}

A low-dimensional system is one where the motion of microscopic 
degrees-of-freedom, such as electrons, phonons, or photons, is 
restricted from exploring the full three dimensions of our world. 
There has been tremendous interest in low-dimensional quantum 
systems during the past twenty years, fueled by a constant stream 
of striking discoveries and also by the potential for, and 
realization of, new state-of-the-art electronic device architectures.

The paradigm and workhorse of low-dimensional systems is the 
nanometer-scale semiconductor structure, or semiconductor 
``nanostructure,'' which consists of a compositionally varying 
semiconductor alloy engineered at the atomic scale\cite{Butcher etal}.
Traditionally 
one would not include naturally occurring low-dimensional entities 
such as atoms and molecules in the subject of this article, but 
some of the most exciting recent developments in the field have 
involved the use of molecules and even biologically important 
materials and has blurred the boundaries between the subject 
and other physical and life sciences. In addition, there are systems 
of great interest in physics, such as high-temperature superconductors,
where the effects of reduced dimensionality are believed to be 
essential, and these too will be regarded as low dimensional.
Many of the subjects covered here are central to the currently
fashionable fields of nanoscience and 
nanotechnology\cite{Timp,Milburn 1997}.

The study of low-dimensional quantum phenomena has led to entirely 
new fields of research, such as the physics of mesoscopic systems, 
which will be discussed below. And low-dimensional systems have shed 
new light on the difficult questions of how disorder (impurities, 
for example) and electron-electron interaction affect a quantum 
system. In fact, understanding the combined effects of disorder 
and interactions in condensed matter systems is currently a problem 
of enormous interest. 

How are electrons, say, restricted from moving in three dimensions? 
The answer is {\it confinement}. Take, for example, an electron inside
a long wire: The positively charged ions in the wire produce an 
electric field that prevents the electrons from escaping. Often, in 
fact, one can regard the electrons as being subjected to a hard-wall 
potential at the wire's surface. The electronic eigenstates are given 
by a plane wave running along the wire multiplied by a localized 
function in the transverse directions. For a range of low energies the
eigenstates have the same transverse eigenfunction and only the plane 
wave factor changes. This means that motion in those transverse 
directions is ``frozen out,'' leaving only motion along the wire.

This article will provide a very brief introduction to the physics of 
low-dimensional quantum systems. The material should be accessible to 
advanced physics undergraduate students. References to recent review 
articles and books are provided when possible. The fabrication of 
low-dimensional structures is introduced in Section \ref{fabrication}.
In Section \ref{theory} some general features of quantum phenomena in 
low dimensions are discussed. The remainder of the article is devoted 
to particular low-dimensional quantum systems, organized by their 
``dimension.''

\section{Making Low-Dimensional Quantum Structures}
\label{fabrication}

The most common method of fabricating low-dimensional structures is by
``growing'' compositionally graded semiconductor alloys in high-vacuum
molecular-beam epitaxy (MBE) machines. Take, for example, the 
homogeneous alloy ${\rm Al}_c {\rm Ga}_{1-c} {\rm As}$ consisting of 
a periodic array of arsenic atoms together with a fraction $c$ of 
aluminum and $1-c$ of gallium. The special cases of $c=0$ and $c=1$ 
correspond to the crystalline semiconductors ${\rm GaAs}$ and 
${\rm AlAs}$, each with a distinct band structure. The microscopic 
potential produced by the alloy ${\rm Al}_c {\rm Ga}_{1-c} {\rm As}$, 
although not strictly periodic, may be regarded as producing a band 
structure interpolating between that of ${\rm GaAs}$ and ${\rm AlAs}$. In 
particular, the energy gap between the valence and conduction bands 
varies with $c$. 

Structures that confine electrons are made by changing the aluminum 
fraction $c$ during crystal growth, leading to a compositionally 
graded alloy of the form ${\rm Al}_{c({\bf r})} {\rm Ga}_{1-c({\bf r})
}{\rm As}$, where $c$ varies spatially. The resulting band structure 
variation produces a spatially varying conduction band minimum. Hence,
an electron added to the conduction band through doping, optical 
excitation, or electrical injection, sees a position-dependent 
potential. By varying $c$ appropriately, one can engineer confining 
potentials that restrict electron motion to fewer than three 
dimensions. 

In practice, however, it is possible to vary $c$ in one direction 
only, resulting in, at best, a two-dimensional system. To complete 
the construction of a semiconductor nanostructure it is often 
necessary to follow growth with lithography, the selective etching of 
prepared surfaces. After coating a surface with a protective material,
the ``resist,'' patterns are imprinted on the resist in a photographic
processes with focused light, electron beams, or even atoms. After the
imprinted pattern is removed chemically, the underlying semiconductor 
is etched away, leaving an environment that confines electrons both 
in the MBE-growth direction and laterally (perpendicular to the growth
direction). Another common way to produce lateral confinement is to 
use lithographic techniques to pattern metallic electrodes, or 
``gates,'' on the surface of an crystal (grown by MBE, for example) 
that has electrons confined in a buried two-dimensional layer parallel
to that surface. By applying voltages to these electrodes, electrons 
in the layer can be depleted from or attracted to the regions below 
the electrodes.

Finally, it is in many cases necessary to attach electrical contacts 
to the electron gases inside these nanostructures and to the metal 
gates on their surfaces.

The lower limit to the size of the structures one can make is usually 
determined by the size of the patterns one can make with lithography, 
which, in turn, is usually determined by the quality of the image 
formed during the exposure stage. At the time of this writing the 
semiconductor industry can produce, in an integrated circuit, wires 
with a thickness of only 10 nm, more than 10 times smaller that the 
wavelength of visible light.

To make even smaller structures naturally occurring nanometer-scale 
systems have used. Examples include self-assembled nanostructures and
structures incorporating molecules or biological materials. A 
self-assembly technique to make arrays of few-nanometer quantum dots 
(small ``artificial atoms'' where electrons are confined in all three 
dimensions) by growing a thin layer of highly strained material on top
of another crystal has been particularly successful. The strained 
layer relaxes by breaking into small islands, which form the quantum 
dots. An excellent example of the use of molecules to make 
low-dimensional systems is provided by large carbon sheets that can be
rolled into stable hollow spheres, known as Bucky balls, or into 
carbon nanotubes. It is even possible to put other molecules inside a 
Bucky ball, to arrange Bucky balls into a molecular crystal, and to 
electrically contact individual nanotubes. Electrical contacts have 
also been attached to other molecules, turning them into 
``transistors.'' 
Polymers and even DNA strands have been used to engineer quantum 
structures at the nanometer scale.

\section{Physics in Quantum Systems of Reduced Dimensions}
\label{theory}

Physics in low-dimensional systems is often different than in three
dimensions. We now discuss some of these differences and the theoretical
tools used to understand them.

\subsection{Effective Mass Theory}

Electrons in semiconductor nanostructures move in the presence of as 
many as three fields; the periodic or nearly periodic potentials 
produced by the atoms in the crystal, fields applied externally, and 
the electron-electron interaction potential. The atomic potentials, 
which vary at the few Angstrom scale, are usually varying much more 
rapidly than the others. In this case, assuming the electronic states 
have energies near the bottom of the conduction band or near the top 
of the valence band, there is an extremely useful description whereby 
the original problem of an electron moving in the presence of the 
atomic as well as other potentials is replaced by the much simpler 
problem 
of an electron, now with a different mass, moving in the presence of 
the slowly varying fields only\cite{Ziman}. For example, the 
effective 
mass of an electron in a GaAs conduction band is about $0.067$ times 
the ordinary bare mass. In general, the effective mass depends on the 
electron's propagation direction, and can even vary with position.

\subsection{Density of States}

An important distinction between systems with different dimensionality
is their density of states $N(\epsilon)$, which is the number of 
states per unit ``volume'' $L^D$ in an energy range $\epsilon$ to 
$\epsilon + d\epsilon$, divided by $d\epsilon$. 
$L$ is the linear size of the system, and $D$ 
can be either 1, 2, or 3. (The zero-dimensional case has to be treated 
separately). A simple general formula can be derived to determine the
energy dependence of $N(\epsilon)$: Given an excitation (electron, 
phonon, photon) with disperion relation $\epsilon(k) \propto 
|k|^\alpha$, the number of states contained within a $D$-dimensional 
sphere of radius $k$ in momentum space is proportional to $(L/2 \pi)^D
k^D$ and the number per unit volume is therefore proportional to $k^D$
or $\epsilon^{D/\alpha}$. The density of states is evidently the 
derivative of the latter with respect to energy, so $N \propto 
\epsilon^{{D \over \alpha}-1}$. Phonons and photons $(\alpha = 1)$ in 
three dimensions have $N \propto \epsilon^2$, and in lower dimensions 
would have $N \propto \epsilon^{D-1}$ as long as the dispersion 
remains linear. Electrons $(\alpha = 2)$ have a density of states 
proportional to $\epsilon^{1\over 2}$, $\epsilon^0$ (i.e., energy 
independent), and $\epsilon^{-{1\over 2}}$ in 3, 2, and 1 dimensions, 
respectively. The expressions for $D=2$ and $D=1$ assume only one
transverse mode or subband present. When more are present, these 
additional branches simply add to $N(\epsilon)$.

%\subsection{Symmetry Breaking and Phase Transitions}
% Mermin-Wagner Theorem
% Lieb
% Kosterlitz-Thouless transition
% classical versus quantum phase transitions

\subsection{Mesoscopic Physics}

Mesoscopic physics is an exciting new field of science made possible 
by nanostructures\cite{Imry}. A mesoscopic system is one that is in 
some sense between a microscopic and macroscopic system. It is 
typically much larger than a few atoms or molecules, yet it is small 
enough that the degrees-of-freedom (usually electrons) have to be 
regarded as fully quantum-mechanical. More precisely, a mesoscopic 
system has a size $L$ that is larger than a microscopic lengthscale 
$a$ (for example the Bohr radius), yet smaller than the 
phase-coherence length $L_\varphi$, 
which is the characteristic length beyond which a particle loses 
phase coherence. $L_\varphi$ generally depends on the dimension and 
the temperature of the system, as well as on microscopic details. Thus,
in simple terms, a mesoscopic system is one that is larger than 
microscopic and in which quantum mechanics manifests itself fully.

In what follows we shall discuss a few examples and concepts from
mesoscopic physics.

\subsubsection{Aharonov-Bohm Effect}

In 1959 Yakir Aharonov and David Bohm predicted that a magnetic field 
can infleuence the quantum-mechanical phase of a charged particles, 
even if the particles are prevented from entering the region 
containing the magnetic field. This proposal, although a direct 
consequence of quantum mechanics, violated everyone's intuition and 
was extremely controversial at the time. The Aharonov-Bohm effect has 
been observed hundreds of times and shows up everywhere in the study 
of mesoscopic systems. Excellent accounts of it are given in quantum 
mechanics textbooks\cite{Sakurai} and review articles\cite{Olariu and 
Popescu,Washburn}.

The basic idea is that a charged particle moving around a closed loop 
of circumference $L$ accumulates, along with its kinematical phase 
$2 \pi L/\lambda$, where $\lambda$ is the deBroglie wavelength of the 
particle, an additonal Aharonov-Bohm phase given by $2 \pi \Phi / 
\Phi_0$, where $\Phi$ is the magnetic flux enclosed by the ring and 
$\Phi_0 \equiv hc/e$ is the quantum of magnetic flux. The 
Aharonov-Bohm phase changes the energies of charged particles and also 
can shift quantum interference patterns. But the effect can only be 
observed if the particles are sufficiently phase coherent.

\subsubsection{Persistent Currents}

A simple example of a mesoscopic effect is as follows: Take a small 
metal or semiconductor ring of circumference $L$. The mesoscopic ring 
can be made using the lithographic techniques described above. The ring
is an ordinary conductor, not a superconductor. It seems reasonable and
is usually the case that in the absence of a magnetic field the ground 
state of the ring is non-current-carrying. However, if the 
phase-coherence length $L_\varphi$ is larger than $L$ then by threading
the ring with a magnetic flux $\Phi$ the ground state (or 
finite-temperature equilibrium state) becomes a current-carrying state,
and current will flow around the ring without ever dissipating, even in
the presence of disorder. The electrons in the ring are affected by 
the magnetic field even though they are never in contact with it, a 
consequence of the Aharonov-Bohm effect.

Nondissipative current-carrying states can occur in a superconducting 
ring too, but in the superconducting case the current-carrying state is
not, strictly speaking, an equilibrium state, but is instead an 
extremely long-lived metastable nonequilibrium state.

As as the ring becomes larger, the magnitude of the persistent current 
decreases. One reason is that when $L$ exceeds $L_\varphi$ the 
electrons no longer are moving coherently around the ring and the 
Aharonov-Bohm effect is no longer operative. The second reason is that 
even if is  $L_\varphi$ was always larger than $L$, the magnitude of 
the maximum persistent current decreases as $1/L$.

\subsubsection{Phase-Coherent Transport}

Electrons in a mesoscopic conductor (a conductor whose dimensions are 
smaller than $L_\varphi$) move as a wave, not as a particle. The 
behavior is similar, in fact, to an electromagnetic wave propagating
inside a wave guide. This wave-like nature leads to many unusual
physical properties\cite{Beenakker and van Houten}.

Perhaps the most profound is the origin of resistance in mesoscopic
conductors. Ordinarily resistance is caused by {\it inelastic} 
collisions of the current-carrying electrons with disorder (impurities 
and other imperfections), other electrons, and with lattice vibrations
(phonons). In a mesoscopic conductor, however, an electron typically 
travels the entire length of the system without undergoing an inelastic
collision. Thus it might seem that there would be no resistance at all.
But to measure the resistance in a mesoscopic system one has to attach 
electrical contacts or leads to it, which are macroscopic. What occurs,
then, in a mesoscopic conductor, is that the resistance is caused by a
combination of {\it elastic} scattering of electron waves from 
disorder and inelastic scattering in the macroscopic leads, although 
the magnitude of the resistance is determined exclusively by the former.

This phenomena is succinctly described by a formula, originally proposed
by Rolf Landauer in 1957, for the conductance $G$ (reciprocal of the 
resistance) of a mesoscopic system. Landauer's formula is simply
$G = |t|^2 e^2/h$, where $t$ is the quantum transmission amplitude
for an electron to propagate through the system ($|t|^2$ is the 
transmission probability). The ratio $e^2/h$ of fundamental constants
has dimensions of conductance and is about $1/ 26 k\Omega$.
The Landauer formula shows that the conductance of a mesoscopic 
conductor is directly related to the quantum mechanical probability
that an electron can propagate through the system without scattering
elastically. It makes no reference to the strength of inelastic
scattering that actually causes the energy dissipation normally
associated with a resistor.

Another important consequence of the phase-coherent nature of electron 
transport in a mesoscopic system is a phenomenon called weak 
localization. To understand weak localization it is first necessary to 
understand ``ordinary'' Anderson localization (sometimes called strong
localization), named after Phillip Anderson. Anderson localization is
the process whereby the eigenstates of an electron gas in a disordered
environment become spatially localized around impurities, causing the
system to behave as an insulator instead of a conductor. Weak 
localization is a very different process that also increases the
resistance of a disordered conductor (although it is not strong 
enough to turn a conductor into an insulator). 

It comes from a very special quantum interference effect
that occurs in systems with time-reversal symmetry: To find the 
probability $P$ for an electron to propagate from point ${\bf r}$ to
point ${\bf r'}$, one has to add the quantum amplitudes $A_i$ for the 
electron to take all possible paths $i$, and then calculate the
modulus squared, $P = |\sum_i A_i |^2.$ (This expression is a 
consequence of Feynman's path integral formulation of quantum 
mechanics, but one can also view it as a generalization of the double-slit 
interference formula to an infinite number of ``slits.'') The 
cross-terms in this expression are responsible for interference.

Usually when open paths with ${\bf r}$ different from ${\bf r'}$
are considered, the randomness in the $A_i$ wash out any quantum
interference effects. However, there is a special class of paths,
closed paths with ${\bf r}={\bf r'}$, where interference effects can
be important. In systems with time-reversal symmetry (which basically
means that there can be no applied magnetic field) there will always 
be pairs of closed paths and their time-reversed counterparts in the 
above summation that have the same amplitude. The result is that the
probability to go from ${\bf r}$ to ${\bf r}$, in other words the 
probability to {\it go nowhere} is enhanced by quantum interference
effects, and this amounts to a measurable increase in resistance.

\subsubsection{Dephasing by Electron-Electron Interaction}

As explained above, weak localization (and more generally, the 
Aharonov-Bohm effect) occurs when the electron motion is sufficiently 
phase-coherent. This fact can be exploited to actually {\it measure} 
the phase-coherence length $L_\varphi$ or phase-coherence time 
$\tau_\varphi$, the characteristic time beyond which the electron 
becomes decoherent, in an electron system. $L_\varphi$ and $\tau_\varphi$
are simply related to each other and only one needs to be measured.
The resistance increase
due to weak localization depends on the number of closed paths that 
contribute to the summation in $P = |\sum_i A_i |^2$, and a given 
closed path contributes as long as its length $L$ is less than 
$L_\varphi$ (if $L > L_\varphi$ the electron would not have the phase 
coherence necessary to exhibit interference). Thus, the magnitude of
the weak localization effect---which can be determined by ``turning the 
effect off'' by applying a magnetic filed---can be used to infer the 
value of $L_\varphi$ or $\tau_\varphi$. 

At low temperatures the dominant mechanism for dephasing is 
electron-electron scattering: A given electron feels a fluctuating
electric field produced by the other electrons that scrambles its phase
after some time $\tau_\varphi$. Theoretically it is predicted that
the dephasing {\it rate} vanishes at low temperature as 
$\tau_\varphi^{-1} \propto T^\beta$, where $\beta$ is a positive 
exponent, but recently Richard Webb and others have given experimental
evidence for a {\it saturation} of $\tau_\varphi^{-1}$ in the 
$T \rightarrow 0$ limit. The physics of low-temperature dephasing is 
currently a problem of great controversey and interest.   

\subsubsection{Thouless Energy}

Mesoscopic physics research has also led to a profound new discovery
about quantum systems in general. In the 1970's David Thouless and 
collaborators showed that any
quantum system possess an important fundamental energy scale, now 
called the 
Thouless energy $E_{\rm T}$. The Thouless energy is a measure of how 
sensitive the eigenstates in a quantum system are to a change in 
boundary
conditions. Specifically, $E_{\rm T}$ is defined as the energy change
of a state at the Fermi energy caused by a change from periodic to
antiperiodic boundary conditions. $E_{\rm T}$ would be zero in a
system---for example an insulator---with localized eigenstates, because
if the wave functions don't extend to the boundaries their energies
will be independent of boundary conditions. Thouless s howed that
the dimensionless ratio $g \equiv E_{\rm T}/\Delta \epsilon$ of 
$E_{\rm T}$ with the energy level spacing at the Fermi energy, 
$\Delta \epsilon$, determines
whether the system is a conductor $(g>1)$ or an insulator $(g<1)$.
In fact, $g$ is equal to the conductance of the system in units of 
$e^2/h$.

%\subsection{Quantum Chaos}

%\subsection{Fractional Charge and Statistics}

\subsection{Integrable Systems}

An extremely important theoretical aspect of low-dimensional quantum 
systems is that there is a large family of one-dimensional models that 
are exactly solvable or integrable, that is, the exact many-body wave 
functions and energies can be obtained. This is especially fortunate 
because perturbation theory, which is needed to develop approximate 
descriptions of three-dimensional interacting systems, tends to break 
down and become invalid in low dimensions. One of the first quantum 
many-body problems to be solved exactly was the one-dimensional 
spin-$1/2$ Heisenberg model with antiferromagnetic nearest-neighbor 
interaction, solved by Hans Bethe in 1931. Later other important 
quantum many-body problems were solved exactly, including (i) the 
one-dimensional Bose gas with delta-function interaction; (ii) the 
one-dimensional Hubbard model, which describes interacting electrons 
hopping on a lattice; (iii) the Kondo model, which describes 
noninteracting electrons scattering from a localized magnetic impurity,
and (iv) one-dimensional spin-$1/2$ antiferromagnets with long-range 
(inverse square-law) exchange interaction. 

The formidable mathematical techniques required to solve these problems 
exactly is more than compensated for by the rich phenomena exhibited 
by the solutions and by the new concepts they have introduced into 
modern theoretical physics\cite{Mattis,Ha}. For example, the explicit 
solution of a 
one-dimensional model of interacting Fermions, known as the 
Tomonaga-Luttinger model, has excited-state solutions describing 
noninteracting {\it bosons}, and no ``dressed'' electrons or 
quasiparticles as predicted from the general picture of interacting 
Fermi systems known as Fermi liquid theory. Understanding this 
solvable model led in the 1980's to a new general theory of 
one-dimensional systems known as Luttinger liquid theory and to a 
powerful new mathematical technique called bosonization.
   
% spin-charge separation
% fractional statistics
% conformal field theory

\subsection{Fermi Liquid Theory and Beyond}

One of the reoccuring themes in low dimensional electron systems is 
whether or not the system is a Fermi liquid. As mentioned above, this
term refers to a specific theory of interacting Fermi systems originally
proposed by Lev Landau in 1957, for which (along this his pioneering work
on superfluidity) he was awarded the 1962 Nobel prize. The modern 
definition of a Fermi liquid is a system of fermions (usually electrons)
where the effect of electron-electron interactions are sufficiently tame
that they can be accurately described with perturbation theory. A
Fermi liquid is in this sense ``close'' to a noninteracting electron
gas.

In low dimensional systems, especially one-dimensional systems, Fermi
liquid theory and many-body perturbation theory can break down, leading
to what is called a non-Fermi liquid. The paradigm for non-Fermi
liquids is the Lutinger liquid, which will be discussed further below.

\section{Two-Dimensional Quantum Systems}\label{2D systems}

Two-dimensional systems are usually the easiest to make in the laboratory, 
and, as discussed above, are the starting point for most lower-dimensional 
systems.

\subsection{Surface Science}

The study of atomically clean surfaces has emerged as a major area of 
modern condensed matter physics and chemistry\cite{Zangwill}. Topics of 
current interest include the growth of surfaces (for example, by 
molecular-beam epitaxy) the physical and chemical properties of atoms 
adsorbed on surfaces, the physical and electronic structure of the 
surfaces themselves, and the transitions between their many phases as 
external parameters such as temperature, pressure, and coverage are varied.
Most bulk phenomena, such as superconducitvity and magnetism, have surface 
counterparts that are interesting in their own right. Surface science is 
blessed with an abundance of surface-sensitive probes that have permitted 
extremely detailed experimental studies.
 
\subsection{Heterostructures}

Semiconductor heterostructures, which are simply 
atomically abrupt 
interfaces between two different semiconductors, can be thought of as 
building blocks for more elaborate nanostructures\cite{Bastard}. The most
important quantities characterizing them are the conduction band and 
valence band offsets, which determine the height of the potential steps 
or discontinuities, seen by electrons and holes. They can be 
classified into two categories: In type I heterostructures both electrons 
and holes are attracted to the same side of the interface whereas in 
type II they are attracted to opposite sides.

A single undoped heterojunction cannot do much; it has no free charge 
carriers and has a bulk-like DOS. However, by adding donor impurities 
to the high-energy or barrier side of the heterostructure, electrons 
will escape to the low-energy side of the interface, leaving the 
then-ionized impurity centers behind. The dipolar charge distribution 
formed from the electrons and the ionized donors creates an electric 
field that confines the electrons to the interface and bends the 
conduction band (and all bands) into a triangiular shape. If the density 
of electrons is low enough only the lowest transverse mode of the 
heterostructure will be occupied, resulting in a two-dimensional electron 
gas (see below). This technique of modulation doping, a term that applies 
anytime the dopant atoms are placed away from the region where the 
electrons or holes reside, minimizes the scattering of charge carriers 
from the ionized impurities, and leads to systems with exceptionally high
mobility.

\subsection{Quantum Wells and Superlattices}

Putting two heterostructures together makes a quantum well, which, 
when undoped, looks like the one-dimensional square-well potential 
familiar from elementary quantum mechanics. The depth of the
well in the conduction band is equal to the conduction band offset, 
and the same holds for the valence band. The main difference is that 
the motion is unrestricted in the two directions perpendicular to 
the growth direction. Thus, single-particle eigenstates are labelled 
by a transverse mode quantum number for the motion in the growth 
direction, an in-plane wave vector, and a spin projection. The 
transverse modes form overlapping energy bands called subbands, because 
they are bands within the conduction band. As electrons are added to 
the quantum well its shape changes because of the potential produced 
by the carriers themselves, but the properties remain qualitatively 
unchanged. At low enough densities, only the lowest transverse mode 
(or subband) is occupied, and the motion again becomes two-dimensional, 
the motion along the growth direction being completely quenched.

Quantum wells don't have to be square in shape; there is nothing that 
prevents one from varying the semiconductor's composition during growth 
to produce other confining potentials. For example, parabolic quantum 
wells have been made that mimic the fimiliar one-dimensional harmonic 
oscillator.

Putting many quantum wells together by alternating layers of well 
and barrier materials leads to a periodic square-well potential known 
as a superlattice, which is analogous to a common textbook example 
of a one-dimensional ``crystal'' called the Kronig-Penny 
model\cite{Kittel}. The term superlattice refers to the fact that 
there are two different lattices present, the microscopic one 
formed by the atoms and the larger one consisting of the periodic 
array of quantum wells. The periodic square-well potential leads to 
the formation of energy bands inside each original band,
called minibands, for the same reason that the periodic potential 
in a crystalline solid forms the original bands in the first place. 
The energy gaps between the minibands are called minigaps.

\subsection{The Two-Dimensional Electron Gas}

The two-dimensional electron gas has made a tremendous impact on 
modern condensed matter physics, material science, and, to a 
lesser extent, statistical mechanics and quantum field theory. It 
lies at the boundary between three-dimensional electron systems, 
which are generally Fermi liquids, and one-dimensional systems, 
which are not. In fact, where the two-dimensional electron gas stands 
in this regard is not known at present, although most indications 
are that is it a Fermi liquid (or possibly what is known as a 
marginal Fermi liquid). It also lies at another boundary, that 
between metals and insulators. The three-dimensional electron gas 
in the presence of disorder (impurities and defects) can be an 
electrical conductor or insulator, but in the presence of any disorder 
a one-dimensional system is always an insulator. Previously it was 
thought that the two-dimensional electron gas was insulating, but 
recent experiments have shown that at low enough densities it can in 
fact be conducting. This and a few of the many other exciting 
areas of two-dimensional electron gas research are briefly described 
below.

\subsubsection{The Electron Solid}

Electrons in a metal or doped semiconductor are usually described 
as being in a liquid or gas-like phase, reflecting the absence of 
long-range positional order. However, in 1934 Eugene Wigner 
predicted in that an electron gas of low enough density, moving in 
a uniform background of positive charge, should crystallize. The 
reason for this is that at low densities the electron-electron
interaction energy becomes larger than the kinetic energy, and by 
forming a crystal the interaction energy is minimized. In three 
dimensions (and zero magnetic field) the Wigner crystal is predicted 
to form, at zero temperature, at electron number densities less than 
$n_{\rm 3D} = [{4 \over 3} \pi (r_{\rm s} a_{\rm B}^*)^3 ]^{-1}$, 
where $r_{\rm s} \approx 67$. Here $a_{\rm B}^* \equiv \kappa
\hbar^2/m^* e^2$ is the effective Bohr radius in a solid with 
dielectric constant $\kappa$ and effective mass $m^*$. The density 
in ordinary three-dimensional metals is much too high to form a
Wigner crystal. In two-dimensions (again zero magnetic field) the 
critical density is $n_{\rm 2D} = [\pi (r_{\rm s} a_{\rm B}^*)^2 
]^{-1}$, with $r_{\rm s} \approx 37$. In semiconductors
the effective Bohr radius can be much larger than that in a vacuum, 
making the required densities easier to achieve.

A strong magnetic field suppresses the kinetic energy of the 
electrons, enabling a crystalline state at higher density. The 
physics of Wigner crystals in strong fields is a rich and interesting 
subject with a lot of current activity\cite{Fertig}.

% phonons/magnetophonons
% quantum phase transition

%The melting of the crystal proceeds through a fascinating series 
%of stages: Above a critical temperature there is a Kosterlitz-Thouless 
%phase transition to a hexatic phase with short-range positional order 
%and long-range orientational order.
 
\subsubsection{The Metal-Insulator Transition}

As mentioned above, experiments have observed metallic states and 
transitions between metallic and insulating states in low-density 
 high-mobility two-dimensional electron systems\cite{metal-insulator
transition}. This was a
great surprise, and the first reports, made in 1994, were initially 
treated with some skepticism. In 1958 Phillip Anderson showed that 
a three-dimensional electron gas could undergo a ``metal-insulator'' 
transition to an insulating state in the presence of sufficiently 
strong disorder. The insulating state is said to be localized  
because the single-particle wave functions become localized 
in space rather than being extended throughout the solid. Then 
David Thouless and Franz Wegner introduced the idea that the 
metal-insulator transition could be regarded as a quantum phase 
transition and they studied the problem using scaling methods 
developed for the study of critical phenomena. In 1979, Elihu 
Abrahams and collaborators developed what is now called the scaling 
theory of localization, which applies to noninteracting electrons 
and which predicts no metal-insulator transition in two dimensions. 
Experiments in the early 1980 were consistent with that prediction.

There is no consensus yet for what the correct theory of 
two-dimesnsional disordered systems is. However, because 
electron-electron interaction was not incorporated in the 1979 
scaling theory of localization, it is reasonable to believe that 
the metal-insulator transition is driven by interactions, and 
there are good experimental indications of this as well.
 
\subsubsection{The Quantum Hall Effect}

The quantum Hall effect is an extremely rich and active area of 
research that can only be touched on here\cite{Prange and Girvin}. 
There are two effects, called the integral and fractional quantum 
Hall effects. They occur when the two-dimensional electron gas is 
placed in a strong perpendicular magnetic field. The eigenstates 
of noninteracting electrons in such a field forms highly degenerate 
bands known as Landau levels. The number of degenerate states in 
each Landau level is equal to the number of flux quanta in the system 
$BA/\Phi_0$, where $B$ is the field strength and $A$ is the system 
area. The Landau levels are separated in energy by the cyclotron 
energy $\hbar \omega_{\rm c}$, where $\omega_{\rm c} \equiv eB/m^*c$. 
If the magnetic field is strong enough all the electrons in the 
two-dimensional electron gas can be accomodated in the lowest 
Landau level. Is is convenient to measure the electron density 
$n$ in units of the density that fills exactly one Landau level; 
this is called the filling factor $\nu$, which can be written as 
$n h c/eB$. The filling factor can be controlled experimentally 
either by changing the electron density or by changing the magnetic 
field.

When $\nu$ is equal to an integer it is observed (at low temperatures) 
that the transverse or Hall conductance of the electron system is, to 
an extremely high precision, equal to $\nu \, {e^2 \over h}$, 
independent of any microscopic details such as the nature and strength 
of the disorder. (The transverse conductance is the ratio of the 
current through the sample divided by the voltage drop across the sample 
in the direction perpendicular to the current flow). 
This integral quantum Hall effect was 
discovered experimentally in 1980 by Klaus von Klitzing and coworkers, 
for which he was awarded the 1985 Nobel prize in physics. At the same 
time that the transverse conductance is quantized, the ordinary 
longitudinal resistance vanishes. This integral quantum Hall effect can 
be understood to be a consequence of the Landau-level structure of the 
noninteracting spectrum. 

More surprising was the experimental discovery starting in 1982 of states 
with quantized Hall conductance at certain fractional values of $\nu$, 
such as ${1 \over 5}, {2 \over 9}, {3 \over 13}, {3 \over 11}, 
{2 \over 7}, {1 \over 3}, {2 \over 5}, {3 \over 7}, {4 \over 9}, 
{5 \over 9}, {4 \over 7}, {3 \over 5}, {2 \over 3}, {7 \over 9}, 
{4 \over 5},$ and so on. In a brilliant 1983 paper Robert Laughlin 
explained the simplest fractional states ($\nu = 1/3$ and $1/5$) by 
essentially guessing the exact many-body wave function for the ground 
and lowest extited states. His work showed that at these special 
densities the two-dimensional electron system forms a highly correlated 
liquid-like ground state with an energy gap of order $0.01 \, \kappa 
e^2/\ell$, where $\ell \equiv \hbar c / e B$ is the magnetic length. 
This energy gap has the same effect as the energy gap between Landau 
levels in the nonnteracting case and leads to a quantized transverse 
and vanishing longitudinal conductance. Laughlin also predicted that at 
filling factors close to $\nu = 1/q$, with $q$ an odd integer, 
the ground states would consist of his correlated state plus 
quasiparticles or quasiholes having a {\it fractional} charge of 
magnitude $e/q$. Not only are these particles fractionally charged, 
they obey ``fractional'' or anyon statistics, intermeditate between 
Bose-Einstein and Fermi-Dirac statistics. Soon after Laughlin's work, 
Duncan Haldane and Bertrand Halperin were able to explain most of the 
other observed fractions by assuming that the quasiparticles themselves 
form correlated Laughlin states when their densities reach certain 
values.   

Laughlin's fractionally charged quasiparticles are believed to have 
been observed experimentally, first by Vladimir Goldman and Bo Su in 
a 1995 tunneling experiment, and more recently by two groups measuring 
the shot noise (current fluctuations) in the fractional quantum Hall 
effect regime. The proposed fractional statistics has not been directly 
observed yet. Robert Laughlin was awarded the 1998 Nobel prize in 
physics for his profound work in this area.

There will not be room to do justice to the incredible amount of 
important work that has been done on this subject. But it is worth
pointing out the physical reason why the properties of the two-dimensional
electron gas in a strong magnetic field are so unusual and often exotic.
The reason is that electron-electron interactions are, in a certain sense,
much stronger than in an ordinary metal. A dimensionless measure of the
strength of interactions is the ratio of the typical interaction energy,
say $e^2/r_0$, where $r_0$ is the average interparticle distance, to
the typical kinetic energy of electrons at the Fermi energy. In an
ordinary metal the latter would simply be the Fermi energy, but in
quantum Hall effect systems where the electrons are in the lowest Landau
level $(\nu \le 1)$ there is no real kinetic energy, because all electrons
in the lowest Landau level have the {\it same} energy, which, without loss
of generality, can be taken to be zero. In this case the dimensionless
ratio is infinite!
Thus, the quantum Hall effect system is one where many-body effects
are extremely important.

% stripes, double-layers, Skyrmions

\subsubsection{Composite Fermions}

In 1989 Jainendra Jain proposed a theory unifying the integral and 
fractional quantum Hall effects\cite{Jain,Heinonen}. 
The idea is to make an 
exact transformation whereby an even number of fictitious 
magnetic flux tubes, each of strength $hc/e$, are attached to each 
electron. The resulting composite object (electron plus flux tubes) 
obeys Fermi statistics and is called a composite Fermion.
The Fermi statistics in this case is a consequence of the 
underlying Fermi statistics of the bare electron combined 
with the Aharonov-Bohm phase shift produced by the flux tubes; 
had an odd number of flux tubes been mathematically attached the 
resulting object would instead be a composite boson. 

The flux-attanchment transformation is an exact transformation 
of the two-dimensional electron gas problem, and the many-body 
Hamiltonian is now much more complicated that it was originally. 
However, when a mean field approximation is applied to this 
transformed Hamiltonian an integral quantum Hall effect of the 
composite fermions results. This technique provides a method to 
reliably calculate properties of fractional quantum Hall effect 
systems. The physics underlying the fractional quantum Hall effect 
is sufficiently complicated that it cannot be described by a mean 
field theory for the original Hamiltonian. The 
magic of the composite Fermion transformation is that it makes an 
exact transformation of the many-body Hamiltonian to a form where 
simple analysis become adequate.

Even more interesting things happen when the transformation is 
applied to non-quantized Hall states, for example when $\nu = 
{1 \over 2}$. In this case there are two real flux tubes per 
electron in the system. Now two fake flux tubes are attached, 
oriented opposite to the real ones, to each particle. On average, 
then, the fictitious magnetic field cancels the real field, leaving 
no magnetic field left. Composite fermion theory then predicts 
that the two-dimensional electron gas, although in a very strong 
perpendicular magnetic field, should behave as though there were 
{\it no} field at all. This astonishing effect has been observed 
in a series of beautiful experiments.

\subsection{High-Temperature Superconductors}

The phenomena of superconductivity, and the related phenomena of 
superfluidity, are extremely interesting subjects that are covered 
in a number of excellent introductory 
texts\cite{Tilley and Tilley,Tinkham}. Superconductivity was discovered 
in 1911 by Kamerlingh Onnes, who observed an abrupt transition to a 
resistance-free state in mercury near $4 \, K$, and, since then, 
there has been a search for materials that become superconducting 
at higher temperatures. In 1986 a new class of oxides were 
discovered by George Bendorz and Karl M\"uller that have 
superconducting transition temperatures as high as $130 \, K$. 
Bendorz and M\"uller were awarded the 1987 Nobel prize in physics 
for this discovery.

These high-temperature superconductors have weakly connected 
two-dimensional sheets composed of Cu and O, and most theories 
of these materials assert that the physics responsible for the 
superconductivity is a consequence of these CuO$_4$ planes. 
Thus, in this sense, high-temperature superconductors are also 
examples of low dimensional quantum systems.

There is no consensus yet as to the microscopic origin of 
high-temperature superconductivity, and a discussion of the 
competing proposals is beyond the intended level of this article. 
However, a few introductory statements can be made.

First, the superconducting state itself seems to be of the 
same type as in conventional superconductors, where electrons 
(or holes) acquire an attractive interaction causing them to bind 
into bosonic pairs which then Bose condense. This description 
of superconductivity, now called BCS theory, was developed by 
John Bardeen, Leon Cooper, and Robert Schrieffer, for which they 
won the 1972 Nobel prize in physics. Beautiful experiments 
measuring the spontaneous magnetization of ${\rm Y Ba_2 Cu_3 O_7}$ 
rings have determined that the pair wave function has $d$-wave 
symmetry, the same symmetry as a $d$ orbital.

When undoped, the CuO layers are antiferromagnetic insulators. 
Electrons removed from them allow in-plane conduction by holes. 
As a hole moves around the surrounding spins must constantly
readjust themselves to lower their antiferromegnetic exchange 
energy. It turns out that the exchange energy can be kept lower 
if the holes move together. Thus, the background antiferromagnetism
is at least partly responsible to the pairing of the holes.

The non-superconducting state of these novel materials has turned 
out to be at least as unusual as the superconduting state. For 
example, it is not even clear whether the normal states are
Fermi liquids or non-Fermi liquids. Certain common features in 
these materials, for example the temperature dependence of the 
resistivity, have no explanation to date.

\subsection{Two-Dimensional Magnetism}

The discovery of high-temperature superconductors with antiferromagnetic 
CuO planes has also stimulated new interest in two-dimensional magnetism, 
especially in two-dimensional Heisenberg antiferromagnets. 

The Mermin-Wagner theorem, which states that a continuous symmetry (in 
this case, rotational symmetry) cannot be broken in a two-dimensional 
system at finite temperature, requires that magnetism in the 
two-dimensional Heisenberg model exist, if at all, at zero temperature 
only. 

Whether or not there is long-range order at $T=0$ depends on
the parameters in the Hamiltonian, and this leads to an interesting
situation: By varying parameters in the Hamiltonian, the quantum ground
state can change, say, from paramagnetic to antiferromagnetic.
This type of transition is called a {\it quantum} phase transition
because it is analogous to thermodynamic phase transitons between 
phases that occur as the temperature is varied, but here the 
temperature is always zero. 

% At low temperatures antiferromagnetic order exists over a finite domain 
% size $\xi$ that diverges as $T \rightarrow 0$. 

In contrast with the Heisenberg model, a finite-temperature magnetic state 
of the two-dimensional Ising ferromagnet, which does not have continuous 
symmetry, does exist.

The study of low-dimensional quantum magnetism continues to be an exciting 
and active area of research. Sophisticated mathematical and computational 
techniques have been developed to extract their physical 
properties\cite{Auerbach}.

\section{One-Dimensional Quantum Systems}\label{1D systems}

One-dimensional systems possess some of the most exotic phases of 
condensed matter. In these systems, the properties of even weakly 
interacting particles differ dramatically from that of noninteracting 
ones, and the generic conducting state of one-dimensional bosons, 
fermions, or even spins is a Luttinger liquid instead of a Landau 
Fermi-liquid. 

\subsection{Quantum Wires}

Quantum wires are extremely narrow wires where electron motion 
is allowed in 
one direction, along the wire, but confined in the other two 
directions. Most often they are created by putting metallic gates on top
of a two-dimensional electron gas and applying voltages to 
deplete the electron gas underneath. This can produce quantum 
wires of varying length whose width can be controlled during 
an experiment. In fact, varying the width changes the number 
of transverse modes that have energies below the Fermi energy 
and that contribute to the conductance of the wire. The quantum 
wires come complete with electrical contacts, made from the 
underlying two-dimensional electron gas, which allow transport 
measurements to be made.

The Landauer formula, described above, predicts that the conductance 
of such a wire is equal to $2 e^2/h$ times the number of transverse 
modes below the Fermi energy. Thus, as the width of the wire is varied, 
the conductance is expected to be quantized in units of $2 e^2/h$. 
By ``quantized conductance'' one means that the conductance is equal
to $2 e^2/h$ times an integer, and that integer increases as the wire
is made wider.

This has been observed in many pioneering experiments, but recent
experiments (on wires made using the cleaved-edge-overgrowth technique)
show quantization in multiples of approxiately 0.7 times that which is 
expected. These intriguing experiments are not understood at this time.

\subsection{Carbon Nanotubes}

Nature has provided beautiful molecular quantum wires called carbon 
nanotubes, formed by rolling carbon sheets of hexagonal symmetry into 
tubes of varying diameter\cite{Saito etal}. They can have diameters as 
little as a nanometer, and can be produced as isolated single-wall tubes, 
isolated multi-wall tubes, or in bundles or ropes of parallel tubes. 

Both their structural and electronic properties are unusual. They are 
predicted to be the strongest fibers known, and do not break when bent. 
In fact, the tube wall appears to simply straighten out  after the 
bending forces are released. Great progress has been made in obtaining 
a microscopic understanding of the structural properties of carbon 
nanotubes.

Much is also known about the transport properties of nanotubes; however, 
there remains many unanswered questions. The sheets of carbon making up 
the tubes, when undoped, are insulators. When they are rolled into tubes 
they can be metals or insulators, depending on the tube diameter and the 
amount of twist, if any, introduced during their creation. Actually, it 
is surprising that a nanotube, if they are regarded as a one-dimensional 
crystal, can be conducting at all, because one-dimensional metallic 
crystals are usually unstable to the formation of a periodic lattice
distortion or strain that lowers their energy but makes them insulating. 
Their tubular geometry, however, is believed to make them sufficiently 
resistant to longitudinal strain.

Then there is the question of electron-electron interaction effects, 
which are usually very important in one-dimensional systems. Luttinger 
liquids (see below) have been predicted to occur in nanotubes, complicated
by the fact that the conducting channels (for each spin orientation) come 
in pairs. Electron-phonon interaction is also believed to be strong in 
nanotubes, and there is even the possibility of turning them into 
superconductors.

\subsection{The Luttinger Liquid State}

In the 1950's and 1960's Sin-itiro Tomonaga and Jaoquin Luttinger studied
theoretical models of one-dimensional metals (similar to quantum wires)
and found that they possessed properties entirely different than 
three-dimensional metals and different than that expected by Landau's
Fermi liquid theory. In particular, the low-energy eigenstates described
elementary excitations that are bosons, not fermions. These bosons are
basically phonons (or quanta of the sound waves) of the electron gas.

The breakdown of Fermi liquid theory in one dimension is now known to
be the norm, and the generic non-Fermi-liquid state of matter in one 
dimension is called the Luttinger liquid. The simplest type of Luttinger
liquid is characterized by a single dimensionless parameter $g$, which,
roughly speaking, characterizes the degree to which the system deviates
from a Fermi liquid, defined as $g=1$. The value of $g$ is determined by 
the parameters characterizing the electron gas, like its density and
electron-electron interaction strength. 

Luttinger liquids are interesting because of their exotic behavior, including
an unusual conductivity as a function of temperature, voltage, and 
frequency, and their extreme sensitivity to disorder. Experimental efforts
to observe them in the laboratory have been unsuccessful until only very 
recently.

\subsection{Edge States of the Quantum Hall Fluid}

The most obvious place to look for a Luttinger liquid is in a quantum 
wire. This has been attempted for years without success because it has 
not been possible to make clean enough quantum wires and disorder ruins 
the Luttinger liquid state, making it insulating. The quantum Hall state
of a two-dimensional electron gas, however, has provided an alternative.

The electrical current present in an ordinary metal (when connected to a 
battery, for example) is just the sum of the current carried by each 
electron. At the same time, when the metal is in its ground or
equilibrium state there is no net current flowing (assuming
a macroscopic conductor so there is no persistent current). This
means that in thermodynamic equilibrium the currents from all the
electrons present must sum to zero. Accordingly, to describe the
current-carrying state of a metal it is sufficient to keep track of
{\it changes} of the state occupation numbers from their equilibrium
values. 

In the linear-response regime, where the metal is only slightly out
of equilibrium, only the occupation numbers near the Fermi energy
change. This is why transport properties like resistance are referred
to as ``Fermi surface'' properties.

Now, in the quantum Hall effect system, when the conditions are such 
that a quantized Hall conductance is observed, the states at the
Fermi energy are physically located at the edges of the two-dimensional
electron gas. (In the bulk of the system, away from the edges, the Fermi 
energy is in an energy gap---either between Landau levels or in the 
energy gap of many-body origin described by Laughlin.) As in an ordinary
metal, to describe the current-carrying state it is sufficient to
keep track of the states at the Fermi energy, which are called ``edge
states.'' The edge state concept has been extremely helpful in 
understanding strong-field magnetotransport in the quantum Hall effect 
regime and in other low-dimensional electron systems\cite{Buttiker}.

Edge states corresponding to {\it fractional} quantum Hall effect
states are especially interesting, because they are predicted to
be Luttinger liquids. However, there some important distinctions
between these Luttinger liquids (called chiral Luttinger liquids) and
the Luttinger liquids described above: First, the value of $g$ in
a chiral Luttinger liquid is predicted to be simply related to the
filling factor $\nu$ of the electron gas, so it can be determined
rather easily. Second, $g$ can be made to differ appreciably from 
1, making the associated non-Fermi-liquid effects larger. And third,
edge states can be made to be entirely insensitive to 
disorder\cite{Buttiker}.

Luttinger liquid behavior has been observed in quantum Hall effect
systems, but the values of $g$ measured do not generally agree with
the theoretical predictions. This is another interesting unsolved 
problem in the physics of low-dimensional quantum systems.

\subsection{Spin Chains}

Spin chains are one-dimensional lattices of spins. Although they occur 
in nature, embedded in three-dimensional materials such as 
${\rm Mn(HCOO)_2 \cdot 2H_2O}$, ${\rm Cs Ni Cl_3}$, and
${\rm Rb Ni Cl}_3$, they are also of great theoretical interest because 
they exhibit dramatic disorder and quantum fluctuation effects that occur 
in higher dimensions as well. One-dimensional spin systems also have a 
large class of exactly solvable versions and can be efficiently studied 
numerically, giving them certain advantages over higher-dimensional 
magnets.

One of the most interesting quantum-fluctuation effects in spin chains 
was discovered by Duncan Haldane in 1983. In higher-dimensional 
antiferromagnets it is believed that the ground states can be 
antiferromagnetically ordered and support a gapless spin-wave spectrum.  
Although true long-range order is prohibited in one dimension, 
quasi-long-range order, where the spin-spin correlation functions decay 
at large distances as a power-law, is allowed, and it was assumed that 
antiferromagnetic spin chains also had gapless spin-wave excitations. 
What Haldane predicted, and what was later established experimentally 
and by numerical studies, was that this picture was correct for spin 
chains with half-integer spins $({1 \over 2}, {3\over 2}, {5 \over 2}, 
\cdots)$ but that integer spin system $(1, 2, 3, \cdots)$ have finite 
energy gaps.

\section{Zero-Dimensional Quantum Systems}\label{0D systems}

Zero-dimensional systems are confined in all three directions, and, 
as such, have discrete energy spectra (isolated energy levels seperated 
by gaps).

\subsection{Quantum Dots}

Quantum dots are nanometer-scale structures, often made from gated and 
compositionally graded semiconductors, that can hold a few to a few 
thousand electrons\cite{Jacak etal,Woggon}. In many respects they are like 
artificial atoms, the inverse-square-law Coulomb force of the nucleus 
being replaced by a linear, harmonic-oscillator-like force provided by 
the confinement. However, unlike natural atoms, the number of electrons 
in a quantum dot can be precisely controlled and varied during an 
experiment, allowing, in some sense, one to ``sweep'' through the 
periodic table and beyond.

Like other nanostructures, quantum dots are studied with both transport 
measurements and optical spectroscopy, with no magnetic field and in 
large fields. They are interesting both from a fundamental physics 
point-of-view and for their potential applications. Applications include 
transistors that control currents at the single-electron level, highly 
efficient low-power lasers, and even circuit elements for quantum 
computers (see below).

\subsubsection{Coulomb Blockade and the Single Electron Transistor}

A simple yet fascinating effect occurs when one tries to pass current 
through a quantum dot: From a classical point-of-view, transport through 
the dot has to occur via a discrete change of charge of the dot, because 
the total charge of $N$ electrons in the dot is $-Ne$, where $-e$ is the
charge of one electron. But this means that the electrostatic energy
$(Ne)^2/2C$, where $C$ is the capacitance of the quantum dot 
(approximately equal to its diameter), changes by a finite amount,
which, for small enough dots, can be larger than the thermal energy
$k_{\rm B} T$ and the energy supplied by the battery $eV$. Here $k_{\rm B}$
is Boltzmann's constant, $T$ is the temperature, and $V$ is the applied
voltage. In this situation no current can flow, a phenomena called
Coulomb blockade because it originates from Coulomb interaction between
electrons. 

This simple classical picture is not entirely correct, because quantum
mechanically an electron can tunnel from one lead to the other, through 
the quantum dot, violating energy conservation on the dot for a short 
period of time. However, this is a small effect (although interesting
in itself) for a quantum dot very weakly connected to leads.

A novel transistor can be made by adding a third lead to the quantum
dot. Unlike the first two leads, however, the third lead is not connected
to the dot itself, but rather, to a metallic gate near the dot. Varying
the voltage $V_{\rm G}$ on this gate changes the electrostatic energy
of the electrons in the dots, and can be used to lower the Coulomb
blockade barrier.

The transistor works as follows: In the presence of the metallic gate
the classical electrostatic energy (when there are $N$ electrons in the
quantum dot) is $E = N^2e^2/2C - eN V_{\rm G}$. For a given value of
$V_{\rm G}$ the optimum number of electrons on the dot (the number that
would minimize $E$) is $V_{\rm G} C/e$. However, $V_{\rm G} C/e$ is
generally not an integer, so $N$ takes the value of the integer closest
to $V_{\rm G} C/e$. As the gate voltage is continuously varied, then, the
number of electrons in the dot and the number passing through the leads 
changes one at a time. We therefore have a transisitor
operating at the level of single electrons!   

A number of other interesting phenomena have been observed in quantum 
dot systems.
A beautiful example is the recent observation of the Kondo effect, which
is a low-temperature many-body effect where the spins of the conduction
electrons
interact very strongly with the spins of the electrons in the dot,
enhancing the conductance of the dot when there is an odd
number of electrons in it. Because quantum dots are somewhat like atoms, 
several classic atomic physics phenomena have also been observed in them.  

\subsubsection{Quantum Computers}

Quantum dots in the near future may even become circuit elements for a 
revolutionary type of computer.

Quantum computing is a new multidisciplinary subject of great current 
interest to university researchers, and also of great importance to 
government agencies and the information and technology 
industries\cite{Milburn 1998}. The term ``quantum computing'' 
refers to the possibility of building a computer out of elements, 
called quantum logic gates, whose operation exploits the laws of 
quantum mechanics to perform operations prohibited by conventional or 
classical logic gates. There is also a related subject called quantum 
information science in which the quantum properties of matter are
to quantify, store, encode, and communicate information. Recent 
theoretical and experimental breakthroughs in these subjects have 
attracted computer scientists, mathematicians, physicists, and 
chemists\cite{Lo etal}.

Whereas the basic unit of information in a classical computer is a binary
digit or bit, a ``0'' or a ``1'' say, a quantum computer processes 
information in the form of a coherent superposition $|\psi\rangle = 
\alpha |0\rangle + \beta |1\rangle$, called a quantum bit or qubit.
Thus, the elementary component in a quantum computer is the two-level
system familiar from quantum mechanics. 

There are two major efforts in quantum computing research. One emphasis is to 
assume that one has a number of qubits that can be manipulated, and ask
what they can be used for. Are there calculations that a 
quantum computer can do much faster than a classical computer? The answer
is yes: If a quantum computer could be built it could factor large 
numbers that would be essentially imposible to factor with a classical
computer using the best known algorithms. And this assumed impossibility
of factoring large integers is the basis for current encription methods
used by the military and by banks. 

The other emphasis is try to design and build a quantum computer. The
qubit referred to above is of course an abstract two-level system. What
is the best physical realization of the qubit? Certainly one wants a
qubit with a long phase-coherence time $\tau_\phi$ (see above), because
this is the time during which the system remains quantum mechanical and 
during which a quantum computation can be performed. Second, one has to
figure out how to manipulate qubits and have them interact to perform
an actual calculation. Electron spin states in quantum dots have emerged
as excellent candidate qubits.

Quantum computing is very new and it is hard to predict if it will
be successful in the near term. But if a quantum computer is successfully
built, it will be revolutionary.

\subsubsection{Exact Diagonalizaton}

A theoretical technique that is particularly well 
suited to study electrons in quantum
dots is the direct numerical solution of the many-body Schr\"odinger
equation. This is usually done by expanding in a basis of noninteracting
eigenstates, so the computational problem becomes that of finding the
eigenvalues and eigenvectors of huge matrices, often called exact 
diagonalization. It is not truely exact because it involves a truncation
of the original infinite-dimensional Hilbert space, but this can often
be done in a controlled manner, yielding results that are insensitive to that
truncation.

Exact numerical studies of quantum many-body systems are usually impossible
because it is only possible to handle a dozen or so particles. But small
quantum dots can be studied experimentally, making exact diagonalization 
methods extremely useful.

\subsection{Artifical Molecules}

Another new area of research is the study of {\it pairs} of quantum dots, 
which then can be thought of as artificial molecules. Just as in natural
molecules, bonds can be formed between the atoms by electrons shared
between dots. These systems permit controlled studies of (at least some)
basic molecular phenomena. One problem of current interest is the
magnetic properties of artifical molecules.

\subsection{Nanocrystals and Nanoparticles}

There is considerable interest in nanometer-scale crystals containing,
say, from $10^3$ to $10^6$ atoms\cite{Gaponenko}. These can have 
structural, electronic,
and optical properties very different than bulk solids. Spherical or
nearly spherical nanocrystals are sometines called nanoparticles. 

Semiconducting and insulating nanorystals have many similarities to quantum
dots. In fact, the electronic properties are often the same, because in
both cases the electron is confined in all three directions. However the
vibrational properties can be quite different: A quantum dot is not usually
mechanically isolated from a bulk solid, so it supports a continuum of
low-frequency vibrational modes. But an isolated nanocrystal would have
a {\it discrete} vibrational spectrum, 
much like that of a drum head. This means
that any physical property involving phonons will be very different than
that in a macroscopic solid.

Metallic nanocrystals have also been actively investigated. They can be
made out of superconductors, allowing for the study of mesoscopic 
superconductivity. Making nanocrystals out of magnetic materials leads
to fascinating new magnetic properties that may have useful data-storage
applications. Metallic nanoparticles also have unusual optical properties
because they are easily polarizable, and the propagation of an 
electromagnetic wave though a cluster of metallic nanoparticles probes 
the electromagnetic modes of that cluster.

\end{multicols}
\end{document}